\documentclass[twocolumn,preprintnumbers,amsmath,amssymb,superscriptaddress]{revtex4}

\usepackage{wasysym}
\usepackage{epsfig}
\usepackage{graphicx}
\usepackage{dcolumn}
\usepackage{bm}
\usepackage{verbatim}

\newcommand{\om}{\omega}

\newcommand{\be}{\begin{equation}}
\newcommand{\ee}{\end{equation}}
\newcommand{\bea}{\begin{eqnarray}}
\newcommand{\eea}{\end{eqnarray}}

\begin{document}

\preprint{APS/123-QED}

\title{Reply to: Atom gravimeters and the gravitational redshift}
\author{}
\author{Holger M\"uller}
\email{hm@berkeley.edu}
\affiliation{Department of Physics, 366 Le Conte Hall MS 7300, University of California, Berkeley, California 94720, USA}
\affiliation{Lawrence Berkeley National Laboratory, One Cyclotron Road, Berkeley, California 94720, USA}
\author{Achim Peters}
\affiliation{Humboldt-Universit\"at zu Berlin, Hausvogteiplatz 5-7, 10117 Berlin, Germany}
\author{Steven Chu}
\affiliation{US Department of Energy, 1000 Independence Avenue SW, Washington, District of Columbia 20585, USA}
\affiliation{Department of Physics, 366 Le Conte Hall MS 7300, University of California, Berkeley, California 94720, USA}
\affiliation{Lawrence Berkeley National Laboratory, One Cyclotron Road, Berkeley, California 94720, USA}

\date{\today}


\maketitle

{\bf We stand by our result \cite{redshift}: The Comment \cite{comment} revisits an interesting issue that has been known for decades, the relationship between test of the universality of free fall and redshift experiments \cite{Schiff,Dicke,Nordtvedt1975}. However, it arrives at its conclusions by applying the laws of physics that are questioned by redshift experiments; this precludes the existence of measurable signals. Since this issue applies to all classical redshift tests \cite{Vessot,ACES} as well as atom interferometry redshift tests, these experiments are equivalent in all aspects in question.}

We first note that no experiment is sensitive to the absolute value of a potential $U$. When two similar clocks at rest in the laboratory frame are compared in a classical redshift test, their difference  $\Delta \nu/\nu =  \Delta U/c^2$ is given by  $U = gh+\mathcal O(h^2)$ , where $g =\vec \nabla  U$ is the gravitational acceleration in the laboratory frame, $h$ is the clock's separation, $c$ is the velocity of light, and $\mathcal O(h^2)$ indicates terms of order $h^2$ and higher. Therefore, classical redshift tests are sensitive to $g$, not to the absolute value of $U$, just like atom interferometry redshift tests. For classical redshift tests, as for any experiment, all effects of conventional gravity, including the redshift, are eliminated locally inside a freely falling laboratory \cite{Will1993}. If the clock's trajectory is tracked from a fixed ground station, the complete experiment is not in free fall. Also, the redshift is  $\Delta U/c^2$, independent of whether the clocks are compared by exchange of electromagnetic signals or clock transport \cite{Will1993}.

The atom interferometry signal $\phi= \phi_{\rm r}+\phi_{\rm t}+\phi_{\rm i}$ is the sum of contributions of the redshift  $\phi_{\rm r} = kgT^2 = \int (mc^2/\hbar)dt$  to the Compton frequency  $\om_C = mc^2/\hbar$ (where $k$ is the wave number, $T$ the pulse separation time \cite{redshift}, $m$ the rest mass of the atom, and $\hbar$ the reduced Planck constant), time dilation $\phi_{\rm t}$, and the laser-atom interaction  $\phi_{\rm i}$. As is common in classical redshift tests, the time dilation $\phi_{\rm t}$ due to the clocks' motion is compensated for, so that a measurement of  $\phi_{\rm r}$ is obtained. This happens automatically, because $\phi_{\rm t} +\phi_{\rm i} = 0$. By interfering the atomic matter waves from the two paths, we obtain $\phi_{\rm r}$, which contains their proper time difference \cite{Borde}. To compare to Einstein's prediction, we measure $g$ by tracking a falling corner cube's position (apart from its proper time) by interfering light reflected off the corner cube.

As has been pointed out \cite{Dicke,Will1993,Will2006}, the Einstein equivalence principle does not follow from tests of the universality of free fall (UFF) without redshift measurements. The Comment, however, argues that we must use the Lagrangian equation (8) to derive the paths as well as the phases. It thus uses a case where measuring the redshift and testing UFF are equivalent and so the Einstein equivalence principle does follow from tests of UFF alone \cite{redshift}. In this case, the Comment is correct that $\phi_{\rm free} \equiv \phi_{\rm r} +\phi_{\rm t} = 0$, and thus atom interferometers measure the acceleration of free fall (as do all redshift tests in this case). Simultaneously, we are correct that $\phi_{\rm  t} + \phi_{\rm  i} = 0$, showing that they measure the redshift. Using this approach, however, precludes redshift anomalies without corresponding UFF violation. Any anomaly would cancel upon comparing measured and predicted redshifts, as the anomaly would also be contained in $g$. This has been shown explicitly for classical redshift tests \cite{Nordtvedt1975}, and atom interferometers are no different in this respect.

The more general - and more interesting - scenario is that redshift and UFF violations can exist independently; explicit theoretical models for this have been proposed using, for example, non-minimal coupling \cite{Coley,Ni} or gravitoscalar and fields \cite{Nieto}. Then, $\phi_{\rm free}$ is no longer zero, but our  $\phi_{\rm t} +\phi_{\rm i} = 0$ still holds \cite{redshift}, even for arbitrary simultaneous violations of UFF. Given that a complete theory exhibiting redshift anomalies while preserving UFF is not yet known, it would be premature to use a detailed model for such physics. Our analysis consequently allows (but does not require) the acceleration of free fall of the atoms $g'$ and the redshift to be independent. Similar assumptions have been applied to all previous \cite{Vessot} and planned \cite{ACES} redshift experiments.

Atom interferometers and classical experiments both provide valid measurements of the gravitational redshift. Both remain important, because they test the Einstein equivalence principle over complementary parameters like elevation (1,000\,km versus 1\,mm), clock frequencies ($10^9$\,Hz versus $10^{25}$\,Hz), clock mechanisms (mostly hyperfine interactions versus mostly strong interactions), and methods (radio link versus clock transport). Considering the wide range of scenarios for physics beyond the standard model, it is important to probe the redshift on all experimentally accessible scales.


\begin{thebibliography}{99}
\bibitem{redshift} M\"uller, H., Peters, A., and Chu, S. A precision measurement of the gravitational redshift by the interference of matter waves. Nature  {\bf 463,} 926-929 (2010).
\bibitem{comment} Wolf, P. {\em et al.} Using atom gravimeters to measure the redshift. Nature {\bf 467,} E1 (2010).
\bibitem{Schiff} Schiff, L. On experimental tests of the general theory of relativity. Am. J. Phys. {\bf 28,} 340-343 (1960).
\bibitem{Dicke} Dicke, R. H. E\"otv\"os experiment and the gravitational red shift. Am. J. Phys. {\bf 28,} 344-347 (1960).
\bibitem{Nordtvedt1975} Nordtvedt, K. Quantitative relationship between clock gravitational "red-shift" violations and nonuniversality of free-fall rates in nonmetric theories of gravity. Phys. Rev. D {\bf 11,} 245-247 (1975).
\bibitem{Vessot} Vessot, R. F. C. {\em et al.} Test of relativistic gravitation with a space-borne hydrogen maser. Phys. Rev. Lett. {\bf 45,} 2081-2084 (1980).
\bibitem{ACES} Cacciapuoti, L. and Salomon, C. Space clocks and fundamental tests: the ACES experiment. Eur. Phys. J. Spec. Top. {\bf 127,} 57-68 (2009).
\bibitem{Will1993} Will, C. M. Theory and Experiment in Gravitational Physics Chapter 2 (Cambridge University Press, 1993).
\bibitem{Borde}	Bord\'e, Ch. J. 5D optics for atomic clocks and gravito-inertial sensors. Eur. Phys. J. Spec. Top. {\bf 163,} 315-332 (2008).
\bibitem{Will2006}	Will, C. M. The confrontation between general relativity and experiment. Living Rev. Relativity {\bf 9,} 3 (2006); http://relativity.livingreviews.org/Articles/lrr-2006-3/.
\bibitem{Coley} Coley, A. Schiff's conjecture on gravitation. Phys. Rev. Lett. {\bf 49,} 853-855 (1982).
\bibitem{Ni} Ni, W.-T. Equivalence principles and electromagnetism. Phys. Rev. Lett. {\bf 38,} 301-304 (1977).
\bibitem{Nieto} Nieto, M. M. and Goldmann, T. The arguments against "antigravity" and the gravitational acceleration of antimatter. Phys. Rep. {\bf 205,} 221-281 (1991).
\end{thebibliography}
\end{document}